\documentclass[twocolumn,prb,aps,superscriptaddress,longbibliography]{revtex4-1}

\usepackage{graphicx}
\usepackage{xcolor}
\usepackage{amsmath}
\usepackage{amssymb}
\usepackage{marvosym}
\usepackage{txfonts}
\usepackage{bm}
\usepackage{xcolor}
\usepackage{color}
\usepackage{ulem}
\usepackage{makecell}
\usepackage{multirow}
\usepackage[colorlinks=true,urlcolor=blue,anchorcolor=blue,linkcolor=blue,citecolor=blue,breaklinks=true]{hyperref}
 
\pagecolor{white}

\newcommand{\lsco}{La$_{2-x}$Sr$_x$CuO$_4$}
\newcommand{\lscozn}{La$_{2-x}$Sr$_x$Cu$_{1-y}$Zn$_y$}
\newcommand{\tbcod}{Tl$_2$Ba$_2$CuO$_{6+\delta}$}

\newcommand{\ybco}[1]{YBa$_2$Cu$_3$O$_{#1}$}
\newcommand{\bscco}{Bi$_2$Sr$_2$CaCu$_2$O$_{8 + \delta}$}
\newcommand{\tc}{\ensuremath{T_c}}

\newcommand{\dwave}{\mbox{$d$-wave}}
\newcommand{\veck}{\textbf{k}}
\newcommand{\veckP}{\textbf{k}^\prime}
\newcommand{\Vkkp}{\big|V^i_{\veck,\veckP}\big|^2}

\newcommand{\FSavgphi}{_{\mathrm{FS}_{\phi}}}

\newcommand{\nvo}{\ensuremath{n_{\mathrm{V}_\mathrm{O}}}}

\def\k{{\bf k}}
\def\={&=&}

\definecolor{forestgreen(web)}{rgb}{0.13, 0.55, 0.13}

\definecolor{LSCORed}{RGB}{214,39,40}

\definecolor{TlBlue}{RGB}{31, 119, 180}

\definecolor{ForestGreen}{rgb}{0.13, 0.55, 0.13}

\definecolor{Orange}{RGB}{255, 100, 0}

\definecolor{Purple}{RGB}{128,0,128}

\begin{document}

\normalem

\title{Optical conductivity of overdoped cuprates from \emph{ab-initio} out-of-plane impurity potentials}

\author{D.~M.~Broun}
\affiliation{Department of Physics, Simon Fraser University, Burnaby, BC, V5A~1S6, Canada}
\author{H. U. \"Ozdemir}
\affiliation{Department of Physics, Simon Fraser University, Burnaby, BC, V5A~1S6, Canada}
\author{Vivek Mishra}
\affiliation{Kavli Institute for Theoretical Sciences, University of Chinese Academy of Sciences, Beijing 100190, China}
\author{N.~R.~Lee-Hone}
\affiliation{Department of Physics, Simon Fraser University, Burnaby, BC, V5A~1S6, Canada}
% \author{S. Mirabi}
% \affiliation{Department of Physics, Simon Fraser University, Burnaby, BC, V5A~1S6, Canada}
\author{Xiangru Kong}
\affiliation{Center For Nanophase Materials Sciences, Oak Ridge National Laboratory, Oak Ridge, TN 37831, USA}
\author{T. Berlijn}
\affiliation{Center For Nanophase Materials Sciences, Oak Ridge National Laboratory, Oak Ridge, TN 37831, USA}
\author{P.~J. Hirschfeld}
\affiliation{Department of Physics, University of Florida, Gainesville FL 32611}

\begin{abstract}

Dopant impurity potentials determined by \emph{ab-initio} supercell DFT calculations are used to calculate the optical conductivity of overdoped LSCO and Tl-2201 in the superconducting and normal states. Vertex corrections are included, to account for the effect of forward scattering on two-particle properties.  This approach was previously shown to provide good, semiquantitative agreement with measurements of superfluid density in LSCO.
Here we compare calculations of conductivity with measurements of THz conductivity on LSCO  using identical impurity, band, and correlation parameters, and find similarly good correspondence with experiment.  In the process, we delineate the impact of the different disorder mechanisms on single-particle and transport relaxation processes. In particular, we reveal the critical role of apical oxygen vacancies in transport scattering and show that transport relaxation rates in LSCO are significantly reduced when apical oxygen vacancies are annealed out. These considerations are shown to be crucial for understanding the variability of experimental results on overdoped LSCO in samples of nominally identical doping but different types.  Finally, we give predictions for Tl-2201 THz conductivity experiments.

\end{abstract}

\maketitle{}

\section{Introduction}
\label{sec:introduction}

The standard model of superconductivity  in  metals  relies upon the BCS  pairing instability, generalized to include attraction mediated by fluctuations other than lattice phonons.  The superconducting state condenses from a Landau Fermi-liquid normal state, which can be significantly renormalized by interactions, but which nevertheless contains well-defined fermionic quasiparticles as the elementary  excitations. Cuprate high-temperature superconductors continue to present clear challenges to the Landau--BCS paradigm, particularly in the underdoped to optimally doped regime, where the normal state is a strange metal and a host of intertwined orders survive as remnants of the Mott-insulating parent compound.  Nevertheless, a Landau Fermi-liquid description is expected to re-emerge at sufficiently high doping levels, as kinetic energy must eventually dominate in the high density limit.   It is therefore valid and important to test the extent to which the Landau--BCS paradigm can provide a description of cuprate physics, particularly on the overdoped side.

A complication in pursuing this approach is the role of disorder, which acts in nonintuitve ways in a $d$-wave superconductor and can mask some of the clear experimental signatures expected in the clean limit.  To this end, we have been pursuing a program that attempts to accurately incorporate disorder into the calculation of physical properties of overdoped cuprates, so that the Landau--BCS approach can be tested against experiment.  A particularly important set of electrodynamic measurements has been carried out on highly crystalline MBE-grown films of LSCO, in which doping has been controllably tuned across the overdoped regime.  In a nutshell, these experiments  revealed features that at first sight seemed at odds with $d$-wave BCS theory.  Superfluid density, $\rho_s$, displayed a clear linear temperature dependence, a hallmark of clean \mbox{$d$-wave} superconductivity, but simultaneously showed a strong correlation between $T_c$ and zero-temperature superfluid density, something that is usually only associated with the pair-breaking effects of disorder.  THz spectroscopy carried out on the same samples showed that the linear temperature dependence of $\rho_s$ was accompanied by large transport scattering rates, of the order of $2 T_c$, and a large fraction of residual, uncondensed spectral weight in the $T \to 0$ limit.  Both of these observations were unexpected for a clean $d$-wave superconductor.

Our approach to capturing this behavior in LSCO has been based on a semirealistic, tight-binding parameterization of ARPES electronic structure, \cite{Yoshida:2006hw} with Fermi liquid corrections applied at the lowest energies.  In our early work, we employed a simplified model in which the defects were treated as point scatterers, for concreteness and computational simplicity.  It was shown that weak, Born-limit scatterers, representing dopant defects located away from the CuO$_2$ planes, provided an excellent and internally consistent description of a wide range of physical properties, including superfluid density,\cite{Bozovic:2016ei,Lee-Hone:2017} THz optical conductivity,\cite{Mahmood:2017,Lee-Hone:2018} and thermal properties.\cite{LeeHone2020}  In the latter case, the approach was successfully extended to include Tl-2201, again basing the calculations on ARPES-determined electronic structure.\cite{Plate:2005}  Despite the convincing agreement with experiment, a number of concerns emerged in response to that early work, in particular
that the use of the Born limit implied arbitrarily small impurity potentials, inconsistent with the actual dopants in LSCO and \mbox{Tl-2201}.\cite{Mahmood:2017} In addition,  questions were raised about the magnitude of the 
normal-state scattering rate $\Gamma_N$,  which includes the combined effect of the impurity potential and concentration, and was taken essentially as a fit parameter. 
Finally, the weakness of the point scatterer approximation was recognized already in Ref.~\onlinecite{Lee-Hone:2018}, where it was noted that dopant impurity potentials in cuprates must have significant spatial extent, given their location outside of the CuO$_2$ planes.  The importance of extended impurity potentials was also pointed out in Ref.~\onlinecite{Wang2022}, although the details of those calculations have been challenged.\cite{Ozdemir_comment,Wang_reply}

 \begin{figure*}[t]
    \centering
        \includegraphics[width=1.0\linewidth,scale=1.0]{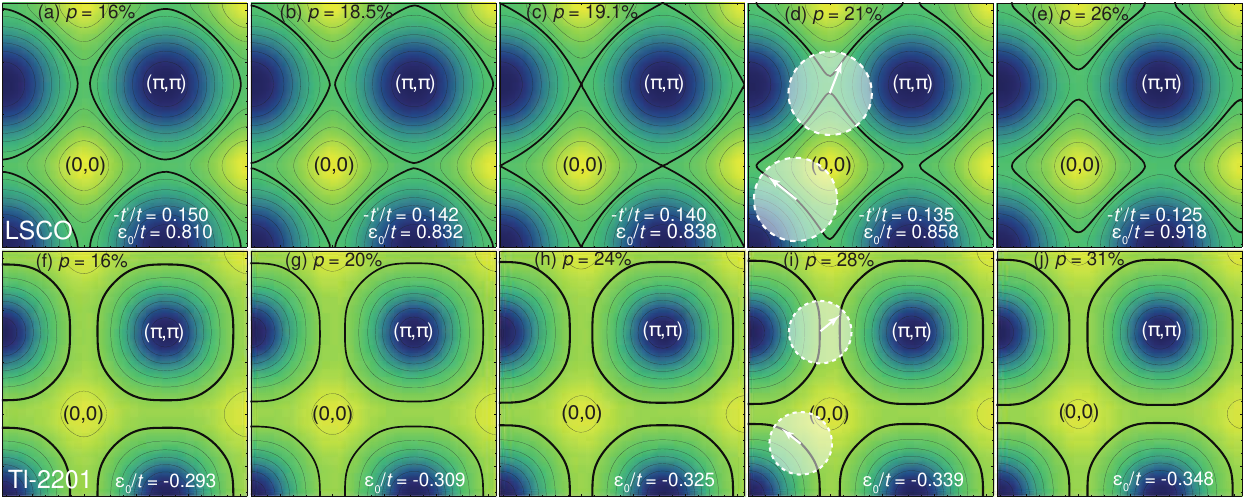}
    \caption{Doping-dependent Fermi surfaces of overdoped LSCO, (a) to (e), and Tl-2201, (f) to (j), showing the Lifshitz transition in LSCO. The Lifshitz transition in Tl-2201 does not occur within the superconducting doping range.  Dashed circles show the FWHM of the central scattering intensities of the Sr dopant in LSCO and the Cu substituent in Tl-2201, respectively, illustrating the effectiveness of umklapp scattering near the antinodes.  Not shown is the scattering intensity for the apical oxygen vacancy in LSCO, which is nearly pointlike and effective at all wavevectors.}
    \label{fig:Fermi_surfaces}
\end{figure*}

To address these concerns, we have embarked on a significant new program, starting with \emph{ab-initio} calculations of the impurity potentials of the three main defect species in LSCO and Tl-2201:\cite{Ozdemir:2022} the Sr dopant and the apical oxygen vacancy in LSCO; and the Cu defect that cross substitutes for Tl in Tl-2201.  These calculations employ a Wannier-function-based approach  to obtain the impurity potentials in tight-binding form, which are then used in subsequent calculations of the dirty $d$-wave superconductor.  As expected, the \emph{ab-initio} potentials are extended in real space, resulting in strongly momentum-dependent matrix elements in \mbox{$q$-space} and requiring vertex corrections for the calculation of two-particle properites.  This procedure was implemented in Ref.~\onlinecite{Ozdemir:2022} for the case of superfluid density.  With the shape and magnitude of the potentials fixed by the first principles calculations, a good, semi-quantitative account of the doping and temperature dependent superfluid density in LSCO was obtained starting only from very reasonable assumptions about defect concentrations.

In the present work, we extend this approach to the calculation of optical conductivity, which is a sensitive probe of the processes that relax charge currents in a $d$-wave superconductor.  We show that with the same impurity potentials, and with assumptions about defect concentration similar to  Ref.~\onlinecite{Ozdemir:2022}, we obtain good, semiquantitative agreement with THz spectroscopy of MBE-grown LSCO thin films,\cite{Mahmood:2017} in terms of the magnitude of the conductivity; the degree of pair breaking and residual spectral weight; and the overall scale of the transport relaxation rate.  The calculations reveal a crucial role for apical oxygen vacancies in transport scattering, due to their scattering potential having nearly pointlike character, resulting in significant scattering intensity at large momentum transfers.  Indeed, within the LSCO system, we can understand the significant differences in residual conductivity/resistivity between ozone-annealed microbridges\cite{Bozovic:2016ei} and cm$^2$ thin films\cite{Mahmood:2017} \emph{entirely} on the basis of apical oxygen vacancies, with the residual resistivity of well-annealed microbridges approaching the intrinsic limit set by the Sr dopants on their own.  Additionally, we provide predictions for Tl-2201, for which no comparable measurements of optical conductivity currently exist.

We begin by briefly summarizing the methods used to capture the electronic structure of LSCO and Tl-2201, followed by sections describing how the \emph{ab-initio} impurity potentials are obtained and then self-consistently incorporated into the theory of disorder pair breaking in a dirty $d$-wave superconductor. Readers interested in further details are referred to our earlier work on superfluid density.\cite{Ozdemir:2022}  We follow this with a presentation of the formalism used to calculate optical conductivity, including vertex corrections.

\section{Homogeneous electronic structure}
\label{sec:formalism}

 \begin{figure}[t]
    \centering
        \includegraphics[width=1.0\columnwidth,scale=1.0]{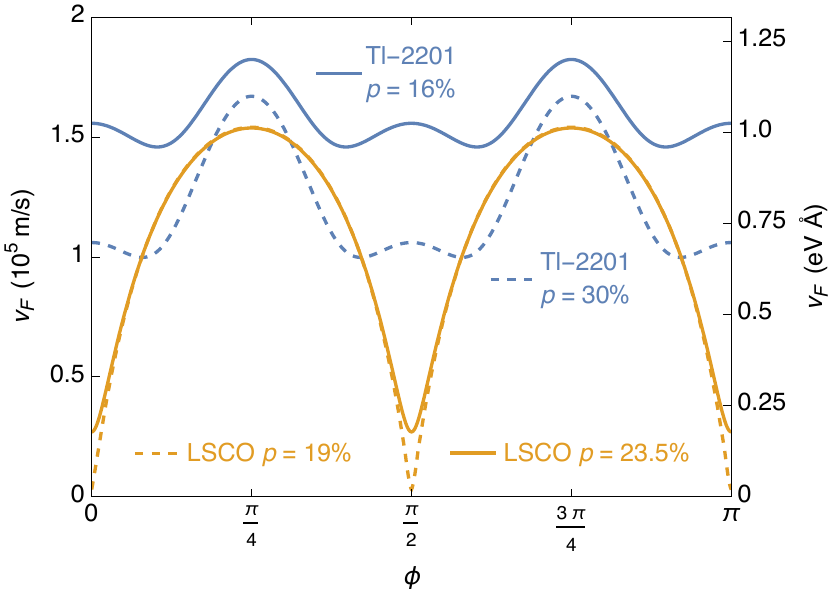}
    \caption{Fermi velocities for LSCO and Tl-2201 as a function of angle around the Fermi surface, measured from the antinodes.  Proximity to the van Hove singularity in LSCO causes deep minima in the antinodal Fermi velocity, even well away from the Lifshitz transition.  Although Tl-2201 is situated far from the Lifshitz transition, a gradual reduction in antinodal velocity is discernible on overdoping.}
    \label{fig:Fermi_velocities}
\end{figure}

As in our previous work on overdoped cuprates,\cite{Lee-Hone:2017,Lee-Hone:2018,LeeHone2020,Ozdemir:2022} our models for LSCO and Tl-2201 are built on two-dimensional tight-binding parameterizations of the ARPES-determined Fermi surfaces and band structures.\cite{Yoshida:2006hw,Plate:2005}  The semi-realistic nature of these models is particularly important for overdoped LSCO, which undergoes a Lifshitz transition around $p = 19\%$ hole doping, as shown in Fig.~\ref{fig:Fermi_surfaces}, at which a van Hove singularity at the antinodal points passes through the Fermi level.  As a result, the electronic dispersion near the antinodes is very flat. This enhances the local density of states and impurity scattering rate near the antinodes but, due to the suppression of Fermi velocity, shown in Fig.~\ref{fig:Fermi_velocities}, makes the contributions from these parts of the Fermi surface relatively unimportant to two-particle, transport-like properties such as superfluid density and optical conductivity.  Nevertheless, it is important that the antinodal regions be treated very carefully: as previously discussed,\cite{Ozdemir:2022} calculations that convert momentum sums to Fermi-surface integrals assuming the usual infinite linearization of the electronic dispersion near the Fermi surface lead to unphysical artifacts such as a strong, doping-dependent enhancement of the overall scattering rate, and therefore of impurity pair breaking, at the Lifshitz transition.  Although computationally more expensive, calculations based directly on momentum sums eliminate these artifacts, and are necessary close to the van Hove crossing.

In the case of LSCO, the doping evolution of the electronic structure is captured by a doping-dependent interpolation of the ARPES tight-binding bandstructures.\cite{Yoshida:2007}  The chemical potential, which is the only parameter with significant doping dependence, is set by the correspondence between hole doping and Fermi volume.  In order to capture the many-body renormalization that occurs at the lowest energies, an overall mass renormalization $m^\ast/m = 2.5$, determined independently via comparison with specific heat data,\cite{Ozdemir:2022, Momono:1994et, Wang2007} is applied to the ARPES bands.

\begin{figure*}[t]
    \centering
   \includegraphics[width=
        0.8\linewidth,scale=1.0]{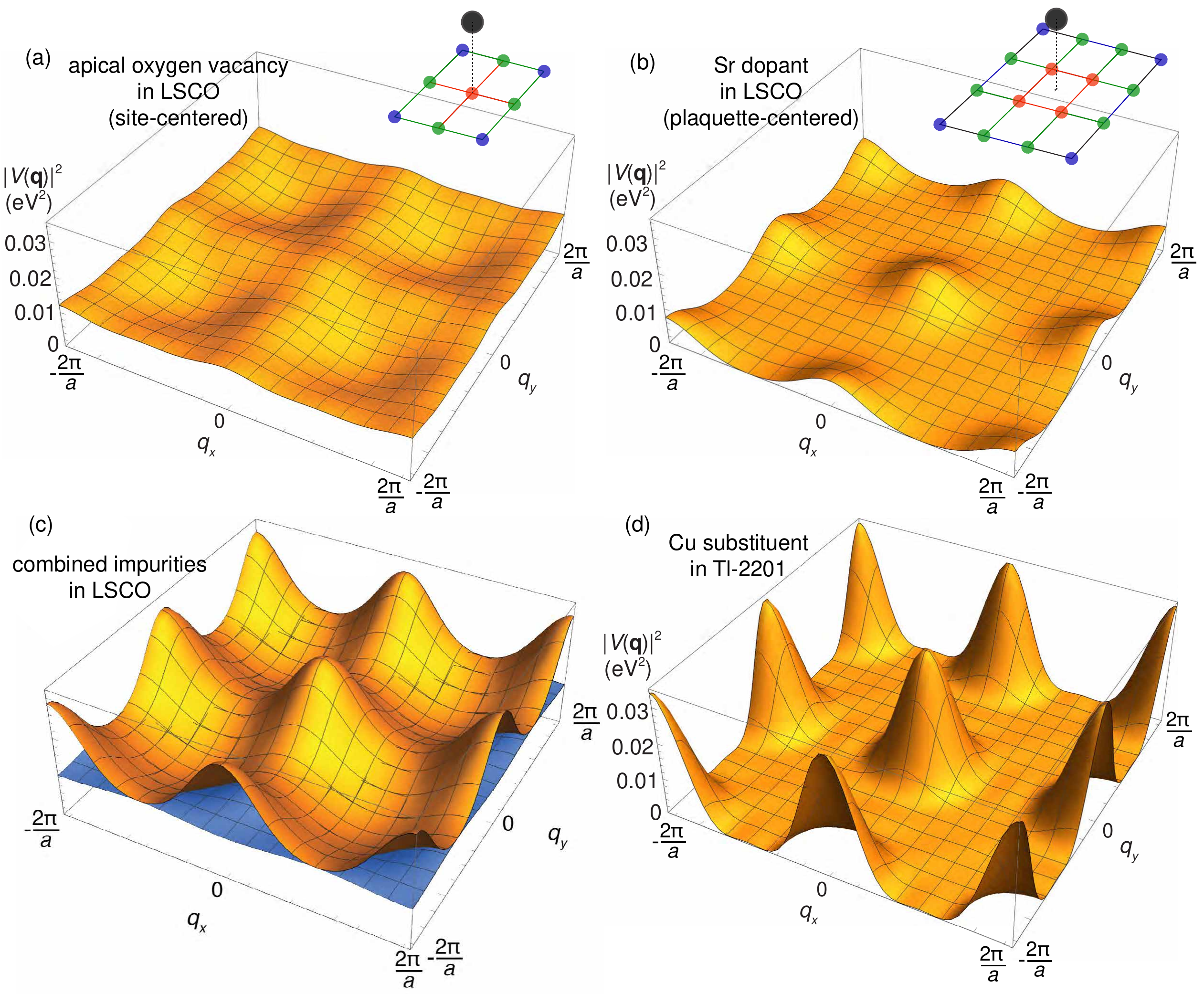}
    \caption{Impurity scattering intensity arising from site-energy terms $V_\mathbf{R}^i$, taken from Ref.~\onlinecite{Ozdemir:2022}, as a function of momentum transfer \mbox{$\veck - \veckP \equiv \mathbf{q} = (q_x,q_y)$} for: (a) the apical oxygen vacancy in LSCO (site centered); (b) the Sr dopant in LSCO (plaquette-centered); and (d) the Cu--Tl substitution in Tl-2201. Panel (c) shows the combined effect of Sr dopants and apical oxygen vacancies in LSCO, for $\nvo = 6$\% and $n_\mathrm{Sr} = 11$\%.
    }
    \label{fig:matrix_elements}
\end{figure*}

The limited ARPES data for Tl-2201\cite{Plate:2005} means that the doping evolution of the Fermi surface is generated via rigid bandshift, with the doping dependence of the chemical potential again set by the Fermi volume.  It should be noted that, unlike for LSCO, there is no additional many-body renormalization required for Tl-2201, as the ARPES measurements of Plat\'e~\emph{et~al.} were carried out at sufficiently low energies (tens of meV) to capture the low energy dispersion directly.

Additional details of the tight-binding dispersions can be found in Refs.~\onlinecite{Lee-Hone:2017} and \onlinecite{LeeHone2020}.  The doping evolution of the LSCO and Tl-2201 Fermi surfaces, shown in Fig.~\ref{fig:Fermi_surfaces}, illustrates the qualitative differences between the materials, and shows how proximity to Fermi-surface replicas in neighbouring Brillouin zones is key to understanding the importance of umklapp processes in antinodal scattering, which are sketched for the Sr dopants and Cu substituents in Figs.~\ref{fig:Fermi_surfaces}(d) and (i), respectively.
    
\section{Impurities in LSCO and Tl-2201}
\label{sec:impurities}

The \emph{ab-initio} calculations of impurity potentials for each of the three main defect species (the Sr dopant and apical oxygen vacancy in LSCO, and the Cu--Tl cross-substitution in Tl-2201) were performed in the following way, as described in more detail in Ref.~\onlinecite{Ozdemir:2022}.  For each type of impurity, two DFT calculations were carried out: one for a $3 \times 3 \times 1$ supercell containing a single impurity (La$_{35}$SrCu$_{18}$O$_{72}$, La$_{36}$Cu$_{18}$O$_{71}$ or Tl$_{35}$Ba$_{36}$Cu$_{19}$O$_{108}$), and one for a pure reference system (La$_4$Cu$_2$O$_8$ or Tl$_4$Ba$_4$Cu$_2$O$_{12}$).  The DFT calcuations were Wannier-projected to define pairs of one-orbital tight-binding Hamiltonians: one for the supercell Hamiltonian for the $i^\mathrm{th}$ impurity,  $H_\mathrm{supercell}^i$, and one for the reference Hamiltonian, $H_0$.   The difference between the two tight-binding models then defines the corresponding impurity potential, in tight-binding form:
\begin{equation}
    \label{eqn:Himp}
    \begin{split}
        H_\mathrm{imp}^i & \equiv \left(H_\mathrm{supercell}^i - \mu^i  {\hat N^i}\right) - \left(H_0 - \mu_0 {\hat N}\right)\\
        & = \sum_{\mathbf{R},\mathbf{R}^\prime\!,\sigma}\!\!\delta H_{\mathbf{R}\mathbf{R}^\prime}^i c_{\mathbf{R}\sigma}^\dagger c_{\mathbf{R}^\prime\sigma}\\
        & \equiv \sum_{\mathbf{R}, \sigma} V_\mathbf{R}^i c_{\mathbf{R},\sigma}^\dagger c_{\mathbf{R}\sigma} + \!\! \sum_{\mathbf{R}\ne\mathbf{R}^\prime\!,\sigma}\!\!\delta t_{\mathbf{R}\mathbf{R}^\prime}^i c_{\mathbf{R}\sigma}^\dagger c_{\mathbf{R}^\prime\sigma}\;,
    \end{split}
    \end{equation}
Here $\mu^i$ and $\mu_0$ are the chemical potentials of the simulation with and without the impurity, respectively, and the 2D lattice vectors $\mathbf{R}$ and $\mathbf{R}^\prime$ are measured in a coordinate system in which the impurity sits directly above (or below) the origin. The impurity potential consists of a set of site energies, $V_\mathbf{R}^i$, along with local modifications to hopping integrals, $\delta t_{\mathbf{R}\mathbf{R}^\prime}$, in the vicinity of the impurity site.  These have been tabulated in Ref.~\onlinecite{Ozdemir:2022}, along with symmetry-generated form factors, and a detailed technical exposition of the DFT calculations and Wannier-projection method. We note  DFT impurity potentials are initially calculated in units of the DFT-derived nearest neighbour hopping $|t|$, with the physically relevant value of $|t|$ subsequently set by taking the experimentally measured value from ARPES, thereby correcting for the tendency of DFT calculations to systematically overestimate the electronic bandwidth in correlated materials.

The real-space impurity Hamiltonian is then recast in momentum space to obtain the matrix elements between Bloch states, $V^i_{\veck,\veckP}$:
\begin{equation}
    \label{eqn:matrixelements}
    \begin{split}
    V^i_{\veck,\veckP} &= \sum_{\mathbf{R},\mathbf{R}^\prime} \delta H_{\mathbf{R}\mathbf{R}^\prime}^i e^{-i \veck \cdot \mathbf{R}} e^{i \veckP \cdot \mathbf{R}^\prime}\\
    &= \sum_\mathbf{R} V_\mathbf{R}^ie^{-i(\veck - \veckP)\cdot \mathbf{R}} + \sum_{\mathbf{R}\ne\mathbf{R}^\prime} \delta t_{\mathbf{R}\mathbf{R}^\prime}^i e^{-i \veck \cdot \mathbf{R}} e^{i \veckP \cdot \mathbf{R}^\prime}.
    \end{split}
\end{equation}
Note that the site energies, $V_\mathbf{R}^i$, give rise to terms that depend only on momentum transfer, $\mathbf{q} \equiv \veckP - \veck$.  (The hopping modifications cannot be expressed in this way, but turn out to be small compared to the site energies.)  This allows the dominant part of the impurity potential to be visualized in $\mathbf{q}$ space.

In Fig.~\ref{fig:matrix_elements}, we plot the scattering intensity
 $|V^i_\mathbf{q}|^2$  for the impurity types that typically occur in LSCO and Tl-2201.\cite{Ozdemir:2022} Due to  the fact that the apical oxygen site located closest to the CuO$_2$ planes is site-centered (see Fig.~\ref{fig:matrix_elements}(a)), the associated impurity potential is nearly point-like, with significant scattering intensity at \emph{all} momentum transfers. Large-$\mathbf{q}$ scattering is crucial to the relaxation of charge currents, particularly in a $d$-wave superconductor, for which inter-nodal scattering dominates electrical relaxation.\cite{Durst:2000}  The Sr dopants, on the other hand, contribute scattering intensity that is concentrated near $\mathbf{q} = 0$ (and umklapp replicas).  This is due, in part, to the Sr site nearest the CuO$_2$ plane being plaquette-centered (see Fig.~\ref{fig:matrix_elements}(b)), which means that a defect at that location affects the four neighboring Cu sites equally, imparting a nonzero range to the impurity potential. 

The doping process in LSCO is often assumed to be synonymous with addition of Sr, i.e., that each Sr simply adds a hole to the band.  However, ARPES measurements on LSCO reveal a discrepancy between Fermi volume and Sr concentration,\cite{Horio:2018} suggesting that the actual relation is \mbox{$n_\mathrm{Sr}=x=0.69p$}. This was considered in Ref.~\onlinecite{Ozdemir:2022}, along with the conventional relation, \mbox{$n_\mathrm{Sr}=x=p$}.  While the assumed form of $n_\mathrm{Sr}(p)$ had no significant effect on superfluid density due to the fact that Sr dopants, with their scattering intensity concentrated near $\mathbf{q} = 0$, are not a strong source of pair breaking, the result highlights the complexity of the doping process. The most likely reason for such discrepancies is the presence of O vacancies in some samples, and a possible negative correlation between the two dopants,\cite{Kim:2017tk} suggesting that high concentrations of Sr dopants drive out apical oxygen.    

The apical oxygen vacancies in LSCO were shown in Ref.~\onlinecite{Ozdemir:2022}  to be the dominant source of pairbreaking if their concentration is significant (at or above the few percent level).  
That they can occur in high-$T_c$ samples to such extent is well-known,\cite{Torrance:1988iz,Higashi1991,Kim:2017tk} but  concentrations are difficult to determine independently. Based on x-ray data,\cite{Higashi1991} even well-annealed crystals (i.e., annealed at 500$^\circ$C for 1~week, in 1~atm~O$_2$) can have apical oxygen vacancies at the 9\% level.  While these results were established decades ago, the O content in the most recent high-quality samples still depends sensitively on geometry and synthesis method, as discussed below.

The plot of $|V^i_\mathbf{q}|^2$ in Fig.~\ref{fig:matrix_elements}(d) reveals why, from a transport perspective, Tl-2201 is qualitatively cleaner than LSCO.  The Tl$_2$O$_2$ double layers, which form an additional structural element not found in LSCO, are relatively well separated from the CuO$_2$ planes. The high volatility of Tl$_2$O$_3$ at the growth temperature leads to a deficit of Tl, which is replaced by Cu on roughly 4\% to 7.5\% of Tl sites.\cite{Liu:1992jx,Kolesnikov:1992gg,Hasegawa:2001bt,Peets:2010p2131} These excess Cu atoms, being further from the CuO$_2$ planes,  produce a softer, longer-range potential than the Sr dopants in LSCO, generating impurity matrix elements that are sharply peaked near $\mathbf{q} = 0$ (and umklapp replicas), with very flat valleys of near-zero scattering intensity in between.   The Cu cross-substituents also play a vital role in the overdoping of Tl-2201, as Cu$^{+}$ has a valence of $-2$ relative to Tl$^{3+}$, making it an effective hole dopant.  As in Ref.~\onlinecite{Ozdemir:2022}, we therefore set the concentration of Cu defects (as a percentage of in-plane Cu atoms) to be $n_\mathrm{Cu} = p/2$, with $n_\mathrm{Cu}$ varying from 8\% to 15\% across the overdoped range. To the extent that it is present, interstitial O$^{2-}$ can be argued to play a similar role, as it similarly dopes two holes, and is located in the Tl$_2$O$_2$ double layers.  This has an additional benefit, as these interstitial oxygen atoms act as an oxygen buffer that minimizes the equilibrium concentration of apical oxygen vacancies.  LSCO has no equivalent oxygen reservoir, so is highly exposed to the formation of apical oxygen vacancies, and the associated strong, point-like scattering potentials.

    \section{Dirty d-wave superconductivity}
    \label{sec:superconducting_state}

The ``dirty $d$-wave'' theory of cuprate superconductors is built around the Nambu-space Green's function of a superconductor.  Within the Matsubara formalism, the renormalized Green's function is written
\begin{equation}
{\underline G}(\veck,i \omega_n)=- \frac{i \tilde\omega_{\veck,n} \tau_0 + \tilde\Delta_{\veck,n} \tau_1 + \xi_\veck \tau_3}{\tilde\omega_{\veck,n}^2 + \tilde\Delta_{\veck,n}^2 + \xi_\veck^2}\;,
\end{equation}
where $\xi_\veck$ is the band dispersion relative to the Fermi level, the $\tau_i$ are the Pauli matrices in particle--hole space, \mbox{$\tilde\omega_{\veck,n}\equiv\omega -\Sigma_0(\k,\omega_n)$} are the renormalized Matsubara frequencies, and $\tilde \Delta_{\veck,n}\equiv \Delta_\k+\Sigma_1(\k,\omega_n)$ is the renormalized superconducting gap.  Note that for the type of momentum-dependent scattering generated by extended impurity potentials, the self energies $\Sigma_0$ and $\Sigma_1$ are both nonzero and are explicitly momentum dependent, unlike the case for point scatterers. In principle, the electronic dispersion is also renormalized as $\tilde{\xi}_\veck = \xi + \Sigma_3$, however, as  argued in Ref.~\onlinecite{Ozdemir:2022}, a $\Sigma_3$ self energy is unnecessary, as any impurity renormalization of the quasiparticle bands is already captured in the phenomenological ARPES-derived dispersions.

The renormalization equations for $\omega_n$ and $\Delta_\mathbf{k}$ take the form
\begin{align}
    &\tilde{\omega}_{\veck,n}=\omega_{n} + \frac{1}{N}\sum_{i,\;\veckP} n_i \Vkkp \!\!\frac{\tilde \omega_{\veckP\!\!,n}}{\tilde \omega_{\veckP\!\!,n}^2+\tilde \Delta_{\veckP\!\!,n}^2+\xi_{\veckP}^2 } -i \frac{\Gamma_N^U}{G_0}   \label{eqn:tau0_self_energy}\\
    &\tilde{\Delta}_{\veck,n} = \Delta_{\veck} + \frac{1}{N}\sum_{i,\;\veckP} n_i \Vkkp \!\! \frac{\tilde \Delta_{\veckP\!\!,n}}{\tilde \omega_{\veckP\!\!,n}^2+\tilde \Delta_{\veckP\!\!,n}^2+\xi_{\veckP}^2}\;.
    \label{eqn:tau1_self_energy}
\end{align}
Here the impurity potentials of the extended, out-of-plane defects, $V^i_{\veck,\veckP}$, are treated to second order, and it was shown in Ref.~\onlinecite{Ozdemir:2022} that the scattering phase shifts associated with these defects are sufficiently weak that the Born approximation is adequate.  We also allow for a small concentration of strong scattering impurities, which are treated as pointlike unitarity scatterers in the $t$-matrix approxation.  These are parameterized by their contribution to the normal-state scattering rate, $\Gamma_N^U$, and generate an additive term in the $\Sigma_0$ self energy, inversely proportional to the momentum-integrated Green's function,  
\begin{equation}
    G_0(i\omega_n) = \frac{1}{\pi N N_0} \sum_\veck \tfrac{1}{2} \mathrm{Tr}\left[\tau_0 {\underline G}(\veck,i \omega_n)\right]\;,
\end{equation}
where $N$ is the number of sites in the lattice and $N_0$ is the DOS per spin at the Fermi level.

As in earlier work based on weak-coupling BCS,\cite{Lee-Hone:2017,Lee-Hone:2018,LeeHone2020,Ozdemir:2022} we assume a separable form for the pairing interaction, $V_0 d_\veck d_{\veckP}$, where $V_0$ parameterizes the pairing strength and the eigenfunction $d_\veck$ takes the form of the simplest $d$-wave harmonic of the square lattice,
\begin{equation}
    d_\veck = \left[\cos(k_x a) - \cos(k_y a)\right]\;.
\end{equation}
Here $a$ is the in-plane lattice parameter, and $d_\veck$ satisfies the normalization condition \mbox{$\frac{1}{N}\sum_{\veck}d_\veck^2 = 1$}.  In terms of this, the \mbox{$d$-wave} BCS gap equation can be written   
\begin{equation}
\Delta_{\veck} = \frac{2 T}{N} \sum_{\omega_{n>0}}^{\Omega_c}\sum_{\veckP} V_0 d_{\veck} d_{\veckP} \frac{\tilde \Delta_{\veckP\!\!,n}}{\tilde \omega_{\veckP\!\!,n}^2+\tilde \Delta_{\veckP\!\!,n}^2+\xi_{\veckP}^2 }\;,
\end{equation}
where $\Omega_c$ is a high frequency cutoff and the $\veckP$ sum runs over the first Brillouin zone.  As shown previously,\cite{Lee-Hone:2017,Ozdemir:2022} the combined effect of pairing strength, $V_0$, and energy cutoff, $\Omega_c$, can be captured by a notional clean-limit transition temperature, $T_{c0}$.  It is important to note that $T_{c0}$ does not imply the transition temperature that would be achieved if disorder was removed from the material. In any real cuprate, inelastic scattering and other fluctuations would destroy superconductivity well before reaching that temperature, something that can be seen, for instance, in the strong downward curvature of $\rho_s(T)$ on the approach to $T_c$ in YBCO,\cite{KAMAL:1994p701} which is not a feature of weak-coupling BCS.    In our model, we take $T_c(p)$ to have the parabolic shape implied by experiment, and solve the gap equation in the presence of disorder to infer $T_{c0}(p)$, with these quantities shown in Ref.~\onlinecite{Ozdemir:2022}, for LSCO and Tl-2201.

\section{Optical conductivity}   
\label{sec:optical_conductivity}
{The electric conductivity can be calculated using the Kubo formula that relates the conductivity, $\sigma$, to the retarded current--current correlation function, $\Pi$:
\begin{equation}
\sigma^{jj}(\Omega) = -e^2 \frac{\mathrm{Im}\,\Pi^{jj}(\mathbf{q}=0,\Omega)}{\Omega},
\label{Eq:Kubo1}
\end{equation}
where $e$ is the electron charge, and $j=x,y,z$ denotes the spatial direction in  real space. Pointlike scatterers do not renormalize the current vertex for even parity superconductors,\cite{Hirschfeld:1988} therefore the bare current--current response is sufficient for calculating conductivity in that case. In contrast,  the bare current vertex is modified in the presence of extended scatterers\cite{Durst:2000} and the current--current correlation function with  impurity-dressed current vertex then reads
\begin{eqnarray}
 \Pi^{jj}(q=0,i\Omega) &=& \frac{T}{N} \sum_{\veck,\omega_n} \mathrm{Tr}\left[ v^j_\veck G(\veck,i\omega_n) G (\veck,i\omega_n+i\Omega) \right. \nonumber \\
 & & \left. \times \,\, \Lambda^j(\veck, i\omega_n,i\omega_n+i\Omega) \right].
 \label{eq:Pi2}
 \end{eqnarray}
 Here $\omega_n$ and $\Omega$ are the fermionic and bosonic Matsubara frequencies, and $v^j_{\veck}$ is the bare current vertex, which for a general dispersion is $v^j_\veck= d\xi_\veck/dk^j$. For the remainder of this section, we replace the full-Brillouin-zone  momentum summation with an angular average over the Fermi surface and integrate out $\xi_\veck$. The initial and final momentum in the impurity potential functions are restricted to the Fermi surface. Since the main contribution to the current--current correlation function comes from quasi-particles near the Fermi surface, this is a reliable and computationally efficient approach for the conductivity in the energy ranges we are interested in. Nevertheless, we have cross-checked the Fermi-surface-based approach against more computationally expensive calculations that employ a full-Brillouin-zone momentum summation and, as long as the Fermi surface is not too close to the van Hove singularity, the two methods are in good agreement.  

\begin{figure*}[t]
    \centering
    \includegraphics[angle=0,width=0.94\linewidth]{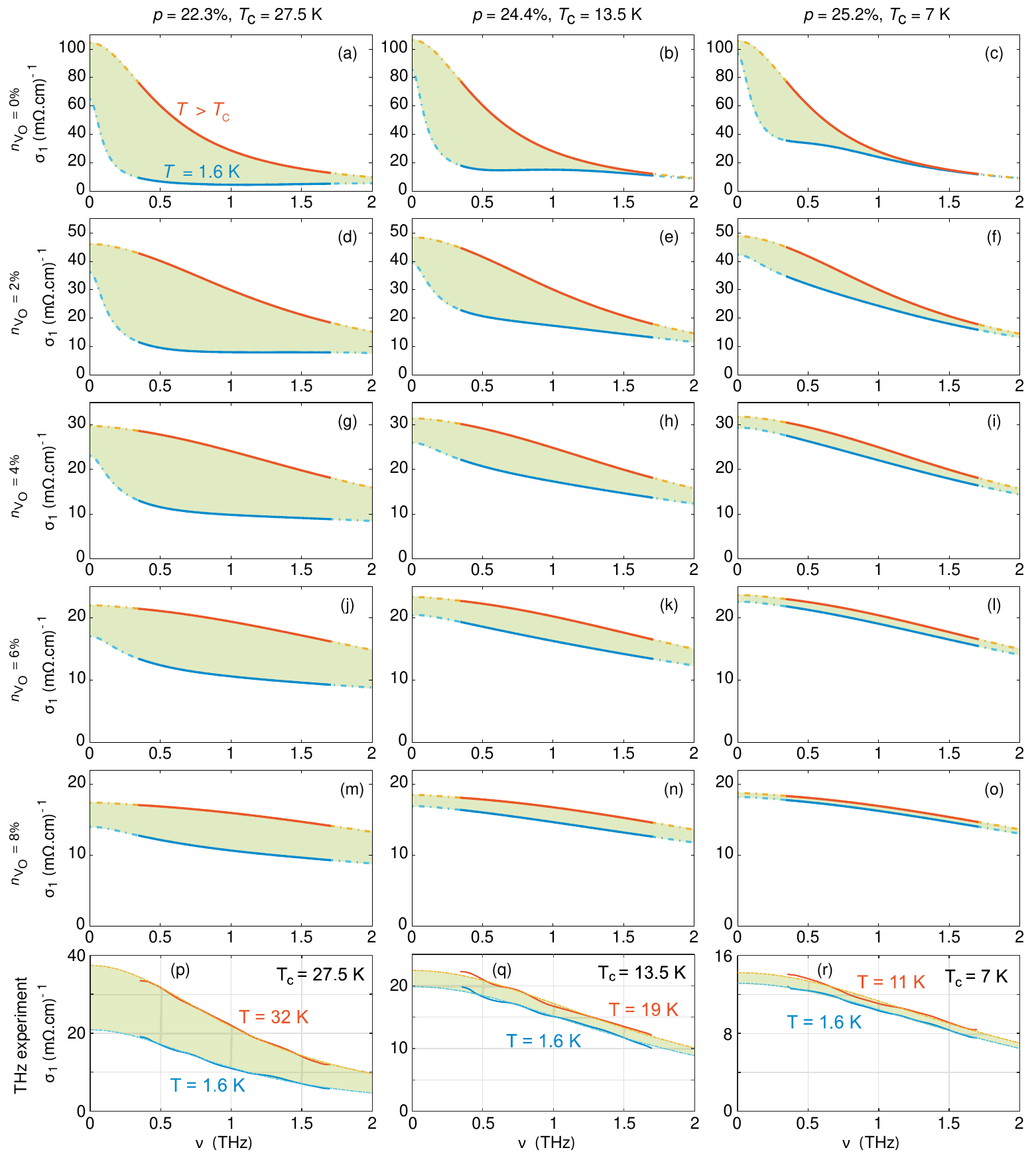}
    \caption{(a) to (o): Calculated optical conductivity of overdoped LSCO, at hole dopings of $p = 22.3$\%, 24.4\% and 25.2\%, chosen to correspond to the superconducting transition temperatures of the samples in the THz study of Ref.~\onlinecite{Mahmood:2017}, reproduced in panels (p) to (r). At each hole doping, the number of Sr dopants is held fixed while the concentration of apical oxygen vacancies is scanned from $\nvo = 0$\% to 8\%, to illustrate separately the effects of the two impurity species.  Each panel contains a normal-state spectrum ($T > T_c$) and a superconducting-state spectrum ($T = 1.6$~K), with the shaded area indicating the spectral weight that condenses to form the superfluid.}
    \label{fig:LSCO_conductivity}
\end{figure*}

\begin{figure*}[t]
    \centering
        \includegraphics[width=1.0\textwidth,scale=1.0]{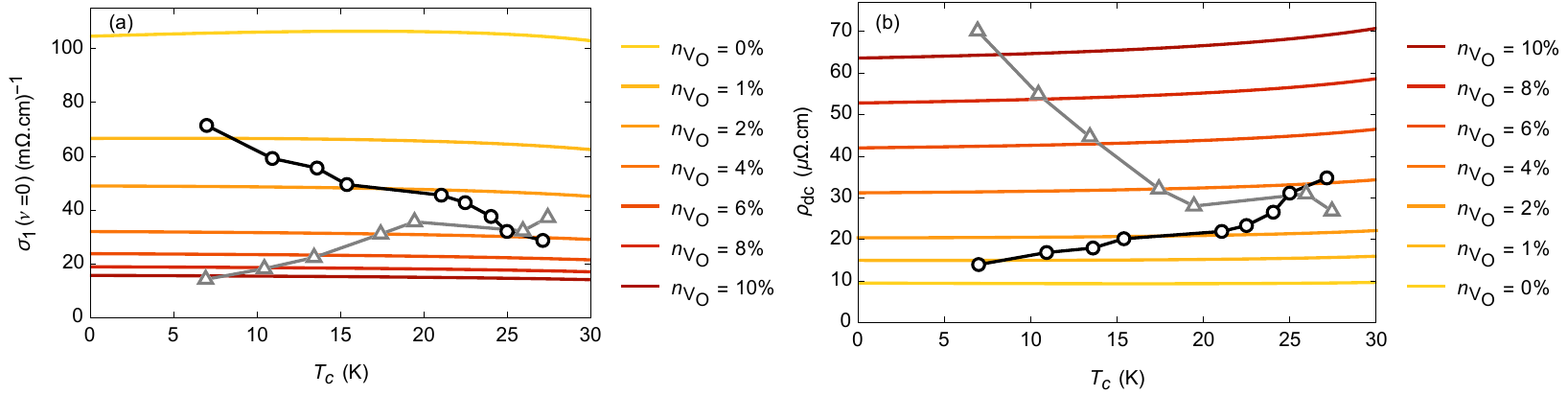}
    \caption{Residual $T \to 0$ normal-state conductivity (a) and resistivity (b) in LSCO,  with theory curves (solid lines) compared to dc transport measurements on ozone-annealed microbridges (open circles) and THz spectroscopy of cm$^2$ thin films (open triangles) from Ref.~\onlinecite{Mahmood:2017}.  Theory curves have been calculated assuming \mbox{$n_\mathrm{Sr} = 0.69 p$} and $\Gamma_N^U = 0.57$~K,  with apical oxygen vacancy concentration ranging from $\nvo = 0\%$ to 10\%.}
    \label{fig:LSCO_residual_conductivity}
\end{figure*}
 
 The vertex correction can be described as
 \begin{eqnarray}
\Lambda^j(\phi,i\omega_n,i\omega_n+i\Omega) &=& v^j_\phi \Big[ \tau_0 \gamma_0(\phi,i\omega_n,i\omega_n+i\Omega) \nonumber \\
& & \left. +\tau_1\gamma_1(\phi,i\omega_n,i\omega_n+i\Omega)\right. \nonumber \\
& &  +\tau_3\gamma_3(\phi,i\omega_n,i\omega_n+i\Omega) \Big],
 \label{Eq:jjcorr}
\end{eqnarray}
where $\phi$ is the angle over the 2D cylindrical Fermi surface.  Finally, for the conductivity we obtain
}
\begin{equation}
    \sigma(\Omega)^{jj} = -\frac{e^2 T}{\Omega} \int_{-\infty}^{\infty} d\omega \big[f(\omega + \Omega) - f(\omega)\big] \left(L^{-+}_{jj} - L^{++}_{jj}\right)\;.
    \label{Eq:sigma}
\end{equation}
The vertex function depends on $i\omega_n$ and $i\omega_n+i\Omega$, and must be  analytically continued to obtain the physical conductivity as a function of real frequency $\Omega$.
In Eq. \eqref{Eq:sigma}, the integrand ${L}^{\pm +}_{jj}$ denotes ${L}^{\pm +}_{jj}(i\omega_n \rightarrow \omega \pm i\eta ,i\omega_n+i\Omega \rightarrow \omega+\Omega  +i\eta)$  and ${L}^{\pm +}_{jj}$ is  
\begin{equation}
    L^{\pm +}_{jj} = \Bigg\langle (v^j_\phi)^2 \left[\gamma_{0,\phi}^{\pm+}(I_{\phi}^{\pm+} + J_{\phi}^{\pm+}) + \gamma_{1,\phi}^{\pm+}K_{\phi}^{\pm+}\right] \Bigg\rangle\FSavgphi\!\!\!\!.
    \label{Eq:Lp}
\end{equation}
Here, the Fermi-surface average is 
\begin{equation}
\left\langle \cdots \right\rangle\FSavgphi = \dfrac{1}{N_0}\int_0^{2\pi} \frac{d\phi}{2\pi} N_\phi \left[ \cdots \right]
\end{equation}
and the angle-dependent single-spin DOS is
\begin{equation}
N_\phi=\frac{|k_F(\phi)|^2}{\pi \hbar d \, v_F(\phi)\cdot k_F(\phi)}\;,   
\end{equation}
where $d$ is the interlayer spacing. The other components of Eq. \eqref{Eq:Lp} are 
\begin{equation}
    \begin{aligned}
        I_{\phi}^{\pm+} & = \dfrac{\tilde\Delta^\pm_{\phi} \tilde\Delta^{\prime+}_{\phi} + \tilde\omega^{\pm}_{\phi}\tilde\omega^{\prime+}_{\phi}}{Q^\pm_{\phi} Q^{\prime+}_{\phi}(Q^{\pm}_{\phi} + Q^{\prime+}_{\phi})} \\
        J_{\phi}^{\pm+} & = \dfrac{1}{(Q^\pm_{\phi} + Q^{\prime+}_{\phi})} \\
        K_{\phi}^{\pm+} & = \dfrac{\tilde\omega^{\pm}_{\phi}\tilde\Delta^{\prime+}_{\phi} +  \tilde\Delta^{\pm}_{\phi}\tilde\omega^{\prime+}_{\phi}}{Q^\pm_{\phi} Q^{\prime+}_{\phi} (Q^{\pm}_{\phi} + Q^{\prime+}_{\phi})} \\
         \tilde\omega_\phi^\pm &= \tilde\omega_\phi(\omega \pm i \delta)\\
        \tilde\omega_\phi^{\prime+} &= \tilde\omega_\phi(\omega + \Omega + i \delta)\\
        \tilde\Delta_\phi^\pm &= \tilde\Delta_\phi(\omega \pm i \delta)\\
        \tilde\Delta_\phi^{\prime+} &= \tilde\Delta_\phi(\omega + \Omega + i \delta)\\
        Q^\pm_{\phi} &= \sqrt{\big(\tilde\Delta^{\pm}_{\phi}\big)^2  - \big(\tilde\omega^{\pm}_\phi\big)^2} \\
        Q^{\prime+}_{\phi} &= \sqrt{\big(\tilde\Delta^{\prime+}_{\phi}\big)^2  - \big(\tilde\omega^{\prime+}_\phi\big)^2}\;.
    \end{aligned}
\end{equation}
Here, the branch cut for the complex square-root function is along the negative real axis.  
The renormalized energy $\tilde \omega(\phi,\omega)$ and gap $\tilde\Delta(\phi,\omega)$  are obtained by solving the self-consistent equations for the self-energies,
\begin{eqnarray}
\tilde{\omega}^\pm(\phi,\omega)&=& \omega \pm i \eta + n_\mathrm{imp} \pi \!\!\int_{\phi'}\!\! \!\! N_{\phi'} |V_{\phi \phi'}|^2 \frac{\tilde{\omega}_{\phi^\prime}^\pm}{Q^\pm_{\phi^\prime}}-\frac{\Gamma^U_N}{g^\pm} , \quad\\
\tilde{\Delta}^\pm(\phi,\omega)&=& \Delta_\phi + n_\mathrm{imp} \pi \!\!\int_{\phi'} \!\! N_{\phi'} |V_{\phi \phi'}|^2 \frac{\tilde{\Delta}_{\phi^\prime}^\pm}{Q_{\phi^\prime}^\pm}.
\end{eqnarray}
Here $g^\pm = \left\langle \tilde \omega_\phi^\pm/Q^\pm_\phi \right\rangle\FSavgphi$,
and the self-consistent equations for the  vertex functions are
\begin{eqnarray}
\gamma_{0 \pm,+} \! &=& \! 1 \! + \!\!\!\int_{\phi'} \!\!\! F_{\phi \phi'} \gamma_{0 \pm,+}\!\left( I_{\pm,+}\!+\! J_{\pm,+}\right) \!+ \!\!\!\int_{\phi'} \!\!\! F_{\phi \phi'}  \gamma_{1 \pm,+} K_{\pm,+} , \quad \\
\gamma_{1 \pm,+}\! &=& \! - \!\!\! \int_{\phi'} \!\!\! F_{\phi \phi'} \gamma_{0 \pm,+} K_{\pm,+}- \!\!\!\int_{\phi'} \!\!\! F_{\phi \phi'} \gamma_{1 \pm,+}\!\left( I_{\pm,+}\! -\! J_{\pm,+}\right) , \quad \\
 \int_{\phi'} F_{\phi \phi'}  &=& \int_{0}^{2\pi}\frac{d\phi'}{2\pi}\pi n_\mathrm{imp} N_{\phi'} |V_{\phi \phi'}|^2\frac{\textbf{v}_{F\phi}\cdot \textbf{v}_{F\phi'}}{|\textbf{v}_{F\phi}|^2}.
 \label{Eq:Vertex_c}
 \end{eqnarray}
 The vertex correction for the $\tau_3$ component vanishes  in this approximation due to particle--hole symmetry near the Fermi surface. The self-consistent equations for the vertex functions are solved using standard iteration methods, followed by numerical calculation of the conductivity. In the next section, we present and discuss the results.
 
\section{Results}
\label{sec:results}

\textit{Ab-initio} calculations of impurity potentials have been presented in Ref.~\onlinecite{Ozdemir:2022}. These potentials were then employed to calculate the optical conductivity of LSCO using the formalism described in Sec. \ref{sec:optical_conductivity}, with the results plotted in Fig.~\ref{fig:LSCO_conductivity}.  In order to compare with the THz experiments from Ref.~\onlinecite{Mahmood:2017} shown in the last row of Fig.~\ref{fig:LSCO_conductivity}, the conductivity calculations were performed for overdoped samples at hole doping levels of $p = 22.3\%, 24.4\%$ and $ 25.2\%$, corresponding to superconducting transition temperatures of 27.5~K, 13.5~K and 7~K.  These doping levels are sufficiently beyond the van Hove doping that the momentum-sum and Fermi-surface-integral methods agree.  For this reason, Fermi-surface integrals have been used to calculate all the conductivities and self-energies presented in this section.  In all cases, the doping-dependent Sr concentration \mbox{$n_\mathrm{Sr}=x=0.69p$} has been assumed but, as with the superfluid density in Ref.~\onlinecite{Ozdemir:2022}, the conductivity spectra are not particularly sensitive to that choice. 

In accordance with the THz experiments, calculations of $\sigma(\nu)$ have been carried out in both the normal state ($T > T_c)$ and deep within the superconducting state ($T = 1.6$~K).   Note that because our model only contains elastic disorder scattering, there is no additional temperature dependence of $\sigma(\nu)$ once we reach the normal state. By virtue of the conductivity sum rule, the shaded regions in between the normal-state and superconducting-state $\sigma(\nu)$ spectra indicate the spectral weight that condenses to form the superfluid density, and therefore provide a graphic illustration of the degree of pair breaking (i.e., superfluid suppression).

To illustrate the importance of apical oxygen vacancies to transport relaxation in LSCO, conductivity spectra are presented for five different apical oxygen vacancy concentrations, ranging from $\nvo = 0\%$ to 8\%.  Comparison with the experimental results from Ref.~\onlinecite{Mahmood:2017} plotted in the bottom row of Fig.~\ref{fig:LSCO_conductivity} show that close agreement with experient is achieved when apical oxygen vacancy concentration is within the range $\nvo = 4\%$ to 6\%.  As pointed out earlier, it is extremely difficult to obtain an independent measurement of apical oxygen vacancy concentration in these materials, but x-ray structural refinements on LSCO single crystals report  $\nvo$ as high as 9\% in well-annealed crystals.\cite{Higashi1991} The THz experiments in Ref.~\onlinecite{Mahmood:2017} were, by necessity, carried out on large, cm$^2$, MBE-grown thin films, and the high degree of crystallinity acheived in the MBE process likely makes the annealing out of oxygen vacancies relatively difficult, due to the need to diffuse oxygen in laterally from the edges of the large samples. 

To further illustrate the sensitivity to oxygen annealing, we show the effect of apical oxygen vacancy concentration on residual ($T \to 0$) normal-state conductivity/resistivity of LSCO, in Fig.~\ref{fig:LSCO_residual_conductivity}.  Here, we compare with two different types of experimental data, taken from Ref.~\onlinecite{Mahmood:2017}, showing dc transport measurements on ozone-annealed microbridges, and THZ spectroscopy of cm$^2$ thin films.  While the data agree at the higher $T_c$ end, they display a striking bifurcation at lower $T_c$ (i.e., when more heavily overdoped), with the ozone-annealed microbridges exhibiting consistently better conductivity/lower residual resistivity.  The experimental data overlay curves of calculated conductivity/resistivity for apical oxygen vacancy concentrations ranging from $\nvo = 0\%$ to 10\%, providing a ready explanation of the variance between the two sample types.  This is consistent with our conjecture that the need to laterally diffuse oxygen in these highly crystalline materials provides a kinetic barrier to annealing out oxygen vacancies in larger samples, with the required diffusion length in the microbridges, by contrast, being only a matter of microns.  For the larger samples, it is also consistent with the %experience 
measurements of Kim et al.,\cite{Kim:2017tk} suggesting that high concentrations of Sr dopants drive out apical oxygen.

\begin{figure*}[th]
    \centering
    \includegraphics[angle=0,width=0.95\linewidth]{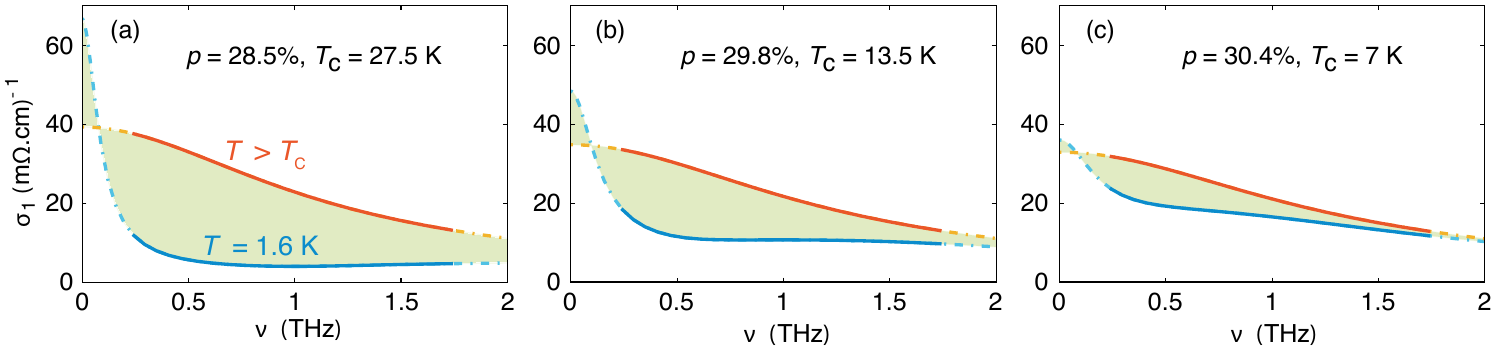}
    \caption{Calculated optical conductivity of overdoped Tl-2201, at hole dopings of $p = 28.5$\%, 29.8\% and 30.4\%, chosen to correspond to the transition temperatures for LSCO shown in Fig.~\ref{fig:LSCO_conductivity}.  As with LSCO, each panel contains a normal-state spectrum ($T > T_c$) and a superconducting-state spectrum ($T = 1.6$~K), with the shaded area indicating the spectral weight that condenses to form the superfluid.}
    \label{fig:Tl2201_conductivity}
\end{figure*} 

We present calculated conductivity spectra for Tl-2201 in Fig.~\ref{fig:Tl2201_conductivity}, with doping levels chosen to give the same $T_c$'s as for LSCO in Fig.~\ref{fig:LSCO_conductivity}.  Due to a lack of suitable Tl-2201 samples, no experimental data on the THz conductivity exist, so these figures serve as a prediction and as a comparison with LSCO. There are several features to note.  While there is some slight doping dependence of the assumed defect concentration (the excess Cu atoms that substitute onto some of the Tl sites) the dominant variation with doping is driven by $T_c$ itself, which in turns sets the size of the superconducting energy gap, and therefore the sensitivity to pair breaking.  For the $T_c = 27.5$~K material, the majority of the spectral weight condenses into the superfluid, leaving a narrow residual Drude peak at the lowest temperatures, riding on a broad background absorption at frequencies out to the gap energy and beyond.  (As previously discussed in Ref.~\onlinecite{Lee-Hone:2018}, in the context of point scatterers, there is usually no sharp gap feature in the optical conductivity of $d$-wave superconductors.)  As $T_c$ becomes smaller (and along with it, the energy gap) the residual Drude peak increases in width and decreases in magnitude, with more and more of the absorption shifting into the broad background.  Interestingly, the calculated spectra for LSCO in Fig.~\ref{fig:LSCO_conductivity} suggest that if cm$^2$ thin films of LSCO could be prepared with apical oxygen vacancy concentrations in the 1\% range, they would show very similar behaviour, i.e., would display charge dynamics that are comparably as `clean' as for Tl-2011, a material often noted for its chemical purity.  This illustrates a somewhat surprising point: that  Tl-2201's reputation as one of the cleaner cuprates is not primarily due to qualitatively lower cation disorder, or to that disorder being located further from the CuO$_2$ planes, but from having an additional structural unit --- the Tl$_2$O$_2$ double layers --- that act as a reservoir for interstitial oxygen, serving as a buffer that suppresses the formation of apical oxygen vacancies.

\begin{figure*}[thb]
    \centering
        \includegraphics[width=0.97\linewidth,scale=1.0]{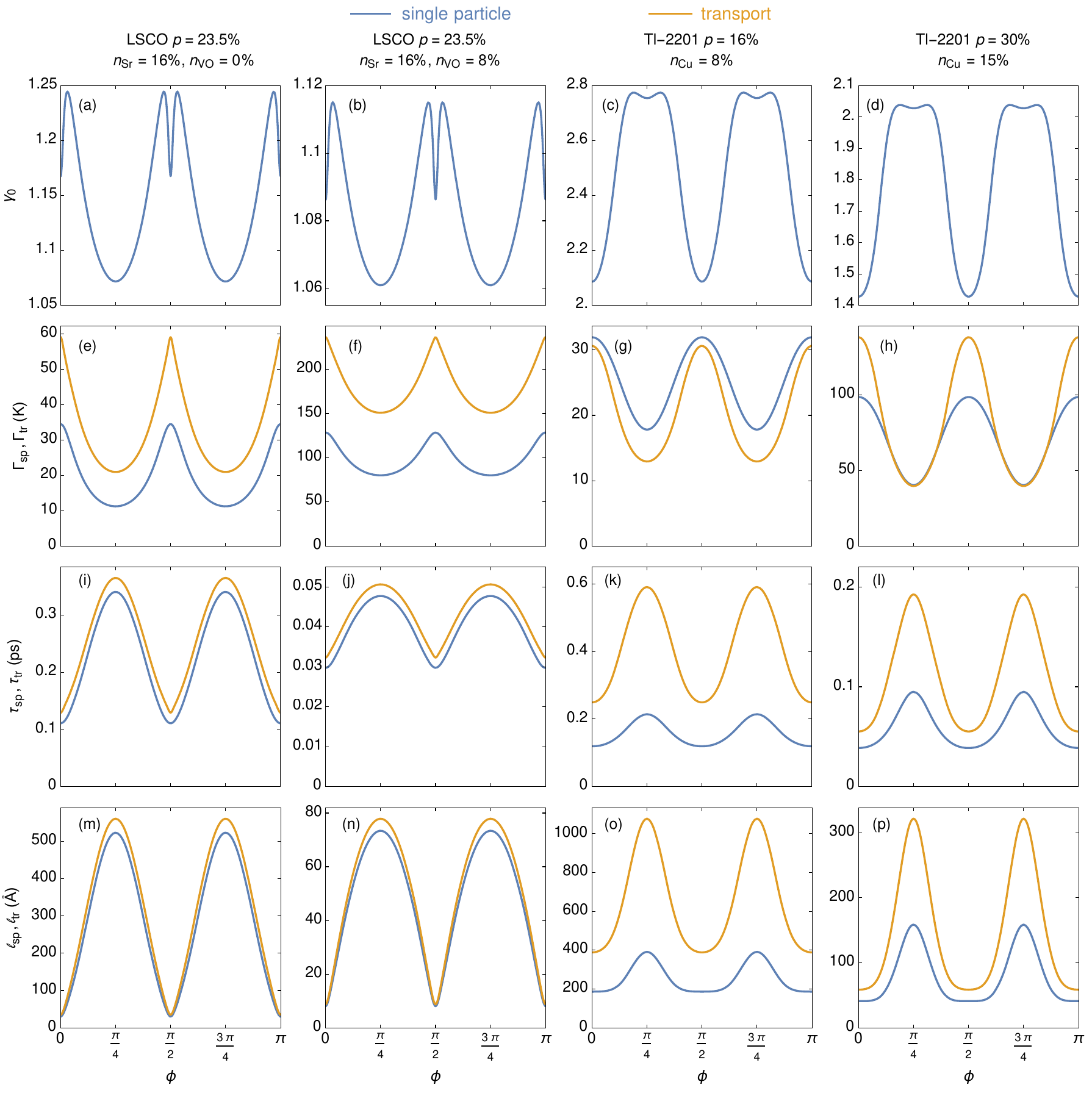}
    \caption{Angle-dependent charge dynamics in the normal state of LSCO and Tl-2201.  Overdoped LSCO, $p = 23.5\%$, with Sr-dopant concentration $n_\mathrm{Sr} = 16\%$ is shown without apical oxygen vacancies (first column) and with apical-oxygen-vacancy concentration $\nvo = 8\%$ (second column).  Tl-2201 is shown at optimal doping, $p = 16\%$, with copper defect concentration $n_\mathrm{Cu} = 8\%$ (third column) and at $p = 30\%$ with $n_\mathrm{Cu} = 15\%$ (fourth column). \mbox{(a) to (d):}  vertex function $\gamma_0(\phi)$, showing enhancement of the charge current due to forward scattering. \mbox{(e) to (h):} single particle scattering rate $\Gamma_\mathrm{sp}(\phi) = \mathrm{Im}\{\Sigma_0(\phi)\}$ and transport scattering rate $\Gamma_\mathrm{tr}(\phi) = 2  \Gamma_\mathrm{sp}(\phi)/\gamma_0(\phi)$.  \mbox{(i) to (l):} Single-particle lifetime, $\tau_\mathrm{sp}(\phi) = 1/(2\Gamma_\mathrm{sp}(\phi))$, and transport lifetime, $\tau_\mathrm{tr}(\phi) = 1/\Gamma_\mathrm{tr}(\phi)$.  \mbox{(m) to (p):} Single-particle mean free path, $\ell_\mathrm{sp} = v_F(\phi) \tau_\mathrm{sp}(\phi)$, and transport mean free path, $\ell_\mathrm{tr} = v_F(\phi) \tau_\mathrm{tr}(\phi)$.\\
    }
    \label{fig:charge_dynamics}
\end{figure*}

To further explore the low frequency charge dynamics of LSCO and Tl-2201 in the normal state, we show angle-resolved plots of scattering rate, lifetime and mean free path in Fig.~\ref{fig:charge_dynamics}. In the case of LSCO, the calculations have been carried out at a fixed doping of $p = 23.5\%$, without ($\nvo = 0\%$) and with ($\nvo = 8\%$) apical oxygen vacancies.  For Tl-2201, we show results for optimal doping ($p = 16\%$) and strong overdoping ($p = 30\%$), with a concomitant change in the concentration of Cu substituents ($n_\mathrm{Cu} = 8\%$ and $15\%$, respectively). A key feature of our calculation is the inclusion of vertex corrections, allowing us to properly take into account the forward-scattering character of the impurity potentials.  This enables us to explore the difference between one-particle and two-particle (transport) scattering
 rates, which differ by the angle-dependent vertex function, $\gamma_0(\phi)$.  This is plotted in the first row of Fig.~\ref{fig:charge_dynamics}, in panels (a) to (d).  We see that vertex corrections in LSCO turn out to be small, even in the absence of apical oxygen vacancies (i.e., when the only scatterer is the Sr dopants, which have a relatively weak scattering potential).  By contrast, in the Tl-2201 system, the vertex corrections lead to significant differences between single-particle and transport lifetimes of order one.

As mentioned above, the spatially extended nature of the realistic disorder model gives rise to impurity matrix elements $V_{\veck,\veckP}$ with very strong momentum dependence. This, combined with anisotropic electronic structure, leads to elastic scattering rates that vary strongly around the Fermi surface, something that is a well-established part of cuprate phenomenology.\cite{Hussey:1996eb,Ioffe:1998p386,Valla:2000en,Abrahams:2000hr,Varma:2001bb,Kaminski:2005ge,Plate:2005,AbdelJawad:2006df,Yamasaki:2007hx,Yoshida:2007,Chang:2008cb,Grissonnanche:2021hw}  Transport is zone-diagonal dominated, as per Ioffe and Millis,\cite{Ioffe:1998p386} due to a combination of factors: the ability for small-$q$ processes to efficiently scatter between antinodes in adjacent Brillouin zones (i.e., to give rise to significant umklapp scattering) and, in the case of LSCO, the deep depression of the antinodal $v_F(\phi)$ in the vicinity of the van Hove singularity.  On the experimental side, a comprehensive Dingle analysis of quantum oscillation data in overdoped Tl-2201 yields single-particle mean free paths in the range 330~\AA\ to 410~\AA, noting that strong self selection in quantum oscillatory experiments preferentially favours the parts of the sample with longest mean free path.\cite{Rourke:2010bl}  This is in qualitative accord with Figs.~\ref{fig:charge_dynamics}(o) and (p).  Two-particle mean free paths inferred from magnetotransport measurements in overdoped Tl-2201 are larger, of the order of 500~\AA\ to 1000~\AA,\cite{Mackenzie:1996p199,Proust:P2lqZi4f,Deepwell:2013uu} confirming both the zone-diagonal-dominated nature of transport, and the presence of vertex corrections of order 2 to 3, in line with our \emph{ab-initio} calculations.

\section{Conclusions}
\label{sec:conclusions}

We have demonstrated that a materials-specific ``dirty \mbox{$d$-wave} approach, previously shown to quantitatively agree with superfluid density data in two of the most-studied overdoped cuprate materials, LSCO and Tl-2201, describes THz conductivity data on the same LSCO films with similar accuracy. Our study has highlighted the role of apical oxygen vacancies in LSCO in  samples produced by different techniques, and  suggested that strong variations in DC resistivity seen in  samples with the same nominal doping are consistent with different levels of O vacancies.  Since the O vacancies produce a relatively large and short-range potential relative to Sr, the O-vacancy concentration has important consequences for the angular dependence of the scattering rate in the normal and superconducting states, and therefore for the relative importance of forward-scattering processes.  Our calculations indicate that if scattering from the O vacancy could be removed by annealing, LSCO would  exhibit dramatically different low-frequency conductivity spectra, with narrow Drude peaks in $\sigma(\Omega)$, reminiscent of our predictions for Tl-2201.  The spectral weight available to form the superfluid would also be significantly increased.

In the Tl-2201 system, the doping by Tl--Cu cross substitution induces much longer-range scattering and relatively weak potentials, leading to a strongly momentum-dependent scattering rate.  Vertex corrections, included here in our calculations of the conductivity, are correspondingly more important.  Although THz measurements of the conductivity have not yet been performed, we make clear predictions for the expected conductivity spectra, including quite narrow Drude components even in the normal state.  The materials-specific analysis confirms  quasiparticle mean free paths that are longer than in LSCO by roughly a factor of three, as deduced in earlier phenomenological analyses.

The materials-specific dirty $d$-wave approach has now succeeded not only in quantitatively reproducing puzzling experimental results on superfluid density and THz conductivity, but also confirmed the choice of phenomenological parameters used earlier to fit specific heat, Volovik effect, and thermal conductivity of the same materials.\cite{LeeHone2020}  Armed with these confirmations of the theory in the superconducting state, it will now be interesting to see if various puzzles in the normal state can be addressed by the same approach, e.g., the angle-dependence of normal-state elastic scattering in cuprates measured by angle-dependent magnetoresistance (ADMR).  Of course, the physics of linear-$T$ resistivity and other non-Fermi liquid effects are not included in this approach, so our theory can perforce only apply in the overdoped regime far from any critical point.  Nevertheless, it will be useful to use it to separate the relatively mundane materials-specific effects discussed from the true exotic physics of interacting fermions located elsewhere in the cuprate phase diagram.

Finally, we should remark that the disorder-averaged theory can also break down when samples become strongly inhomogeneous.  A recent study showed that in quantum simulations of disordered $d$-wave superconductors, the disorder-averaged theory was accurate to surprisingly high disorder levels, but broke down for very low average superfluid densities when the system broke up into distinct islands at low temperatures.\cite{Pal2023}  Such patchiness of isolated superconducting regions has indeed  been observed in some samples of LSCO.\cite{Tranquada2022} Whether the ideal disorder-driven zero-temperature transition to the normal metal is controlled in the best samples by pairbreaking or inhomogeneity is an important open question that requires further experimental work.

\begin{acknowledgments}
We are grateful for useful discussions with N.~P.~Armitage, J.~S.~Dodge, S.~A.~Kivelson, T.~A.~Maier, D.~J.~Scalapino, J.~E.~Sonier, and J.~M. Tranquada.  D.M.B.\  acknowledges financial support from the Natural Science and Engineering Research Council of Canada.  P.J.H.\ acknowledges support from NSF-DMR-1849751.
V.M.\ was supported by NSFC \mbox{Grant No.~11674278} and by the priority program of the Chinese Academy of Sciences \mbox{Grant No.~XDB28000000}. The first-principles calculations in this work (X.K.\ and T.B.) were conducted at the Center for Nanophase Materials Sciences and used resources of the Compute and Data Environment for Science (CADES) at the Oak Ridge National Laboratory, which is supported by the Office of Science of the U.S.\ Department of Energy under Contract No. DE-AC05-00OR22725. In addition we used resources of the National Energy Research Scientific Computing Center, a DOE Office of Science User Facility supported by the Office of Science of the U.S.\ DOE 
under Contract No. DE-AC02-05CH11231.
\end{acknowledgments}

%\bibliography{overdoped}

%

\end{document}